*Article*

# System Energy Assessment (SEA), Defining a Standard Measure of EROI for Energy Businesses as Whole Systems

**Philip F. Henshaw [1]\*, Carey King [2] and Jay Zarnikau [3]**

1 HDS Systems Design Science, Synapse9.com,  New York, NY 10040 USA
2 Center for International Energy and Environmental Policy, Univ. of Texas, Austin, TX 78713 USA
3 LBJ School of Public Affairs & College of Natural Sciences, Univ. of Texas, Austin, TX 78713 USA

E-Mails: *eco@synapse9.com; careyking@mail.utexas.edu; jayz@mail.utexas.edu

* Author to whom correspondence should be addressed; Tel.: +1- 212.795-4844;

**Abstract:** A more objective method for estimating the energy demands for business systems, System Energy Assessment (SEA), is a "Scope 4" energy assessment for purchased energy demands ($E_S$ or PED) as a system environmental cost, the 2010 world average $E_S$ for $1 of GDP being ~1883kWh. Energy needs traceable to business technology are combined with to energy for all other operating services.  Energy is counted for businesses to operate, in working units of matched controlled and operating services, to compose a whole self-managing net-energy system.  That serves to correct a natural blind spot for energy information.  Current methods rely on counting traceable energy receipts. Self-managing services employed by business outsource energy uses to operate, but don't report it. Those untraceable energy demands are often an uncounted 80% of the total embodied energy of business end products.  That scale of "dark energy" in the total is implied by balancing global accounts, so the average energy cost per dollar for businesses matches the world average per dollar of GDP. SEA is a way to get an accurate total energy demand for business products, demonstrated using a detailed case study of a model Wind Farm. The ratio of energy produced to total energy cost is its energy productivity to society ($EROI_S$). Presently outsourced energy needs, paid for from revenues but "unrecorded", are assumed for lack of information to be "0".   Our default assumption is to treat them as "average". The resulting change in embodied energy increases the scale of known environmental impacts, and mitigation costs, attributable to business products.  For example, counting only energy uses of technology gives high values of EROI for labor intensive practices, being assigned no energy cost, as one of the kinds of misinformation that result from not counting the dark energy involved.  Our result partly confirms similar findings by Hall et. al. in 1981 [9] for asking the same general question. We use an exhaustive search for what a business needs to operate as a whole system to find what to

count. That locates a natural physical boundary of the working parts it needs to operate as a unit, defining a business and its measures as physical rather than statistical subjects of science .

## 1. Introduction

### 1.1 Overview

One of the more difficult problems encountered in measuring the energy consumed in producing energy (EROI)[1, 2] is deciding what contributing energy costs to count. It has been a long discussed question whether to include the energy costs of supporting employees along with the fuel uses for various production technologies, as some other ecological economists have also explored [3, 4, 21, 9]. Historically economists have treated technology and business services including labor as independent parts of the economy, calling one "production" and the other "consumption", to define statistical categories for physical accounting of businesses, overlooking their functional connections. The energy demand on the environment became "production costs", counted as the energy receipts traceable to the supply chain for technology used in the business workplace. Other outsourced energy use needs, paid for from business revenues to operate but treated as in other categories, go uncounted. That assumption is presently used for Life Cycle Assessment (LCA) [26] following the world ISO 14000 standards for measuring business energy use, corresponding to green house gas (GHG) "Scope 3" protocol accounting contributions for measuring environmental impacts of business products.

We illustrated our approach using the model business plan for a Texas wind farm from our first study [29]. We assess both the traceable and the untraceable energy needs, based on whether they are required for the business to operate, and compare the totals. We find that counting only the traceable energy uses implies wind energy would be produced with an EROI of 31:1 producing energy at a rather low breakeven Levelized Cost of Electricity (LCOE) of $.002/kWh, using only ~20% of the average energy to produce $1 of value. When counting the total energy demand the, EROI declines to 6:1, there is a 500% increase in the total energy accounted for, LCOE increases to a more realistic $.075-.085/kWh, and a $ of revenue is produced at a more realistic 105% of the world average energy cost.

The method used, "system energy assessment" (SEA), is a physical measure of the total purchased energy demand (PED) for a business, and "Scope 4" energy assessment for total GHG impacts. The important departure we make is using a more objective and comprehensive method of deciding what energy uses to count, tracing physical causation rather than using statistical categories of energy use. It forces us to view both the consumption requirements for machines and for retaining qualified people and other services to operate machines to make businesses work, as equal energy demands on the environment. We count the energy uses of technology in the usual way, relying on the trail of purchase receipts as traceable records. Estimates for the energy demands of the equally essential self-managing services employed are based on the predictable energy needs for the cost of their economic services.

Using SEA, the total purchased energy demand of businesses as whole physical net-energy systems becomes physical measure, called system energy $E_S$ or product energy demand (PED). Then the energy return on business operations $E_R$ is produced with an energy efficiency to society of $E_R/E_S$, to be also called $EROI_S$ for "societal" or "system" energy return on energy invested. We dispense with these suffixes when the intent is clear in context. What makes $E_S$ a physical measure is the exhaustive search for energy needs for the whole working system it measures, defining the physical system with its energy needs boundary. Values of $EROI_S$ as the total energy costs of energy for different businesses, technologies and societies are, are then comparable like any other well defined physical measure such as heat or weight. Measures of $EROI_T$ (for production technology alone) omit the large amounts of untraceable "dark energy" outsourced for business services, resulting in an inaccurate measure only comparable for similar technologies in isolation from the businesses and societies using them. $EROI_T$ for labor intensive energy production, for example, would appear high compared to energy obtained with sophisticated technology, just because the energy costs for labor are not counted.

*1.2 Scientific methods*

The SEA method arose as a special application of a more general "total environmental assessment" method (TEA) [10] designed to identify and anticipate change in the organized complexity of natural systems that operate as self-organizing units. Such systems generally include both passive parts and active processes, that develop and subside as they use their environments. That makes TEA and SEA studies of self-organizing systems rather than of deterministic ones, using an empirical rather than abstract modeling approach. As life cycle assessment, TEA focuses on the normally expected succession of changes in direction in the development of complex energy using systems, from inception through growth, responding to limits and eventual decline. SEA reduces that to measuring economic energy use for businesses as whole systems over a period of time.

Everyone recognizes businesses as having matched active and passive parts organized to work together as a unit. It has not been possible for customary scientific methods to define, measure or refer to them as physical subjects. SEA extends the scientific method to uniquely identify them as units of organization in the environment, using an empirical method arising from complex systems theories [20]. It identifies such systems in their own natural form by identifying them with a reproducible way to define a form fitting boundary and quantitative measures. That expansion of the scientific method for defining complex systems and their measures allows the sciences to treat complex systems as physical subjects, connecting money and energy, so physics can fully apply to economics and the systems sciences such as economics and ecology can broaden the questions of physics.

The procedure for fitting the accounting boundary to the system starts from any part of the business, and tracing physical causations locates everything else the business needs to operate as a whole. We just repeatedly ask: "What else is needed to make it work?" for the business as a self-managing entity to operate in its environment. That provides an objective method for locating the boundary for the working parts as a discovered feature of the business, to the degree of fit that is practical. Adding up all the energy costs for parts within that empirically located boundary results in a quantitative scientific energy measure, and identifies it as a bounded network of working parts. The features of its internal and external relationships can then be studied, as physical subjects with energy budgets of their own, despite having complex parts and features not definable from the information used to locate its boundaries. More discussion of complex systems theory is not needed to use the SEA method.

How one studies a physical system that is more complex than the information that identifies it is like how a tree leaf is revealed by a simple "leaf print" or a broken bone is revealed by an "x-ray". The one kind of information is used to project features of a complex natural system. It reveal useful missing information about it that may be further explored, generally raising good questions by exposing the natural forms for study. That step is what makes this way of accounting for businesses as whole systems a bridge to studying them as a physical science rather than just a statistical science.

Unifying the questions of the sciences around complex systems as objects of the environment allows them to be studied from those multiple perspectives. Science has previously needed to discuss complex systems only in relation to each field's own abstract models. Models of the same complex subject from different views might be different, but at least they would be understood to be connected by referring to the same thing, and not unrelated by being different. Some brief discussion of how to use SEA and $EROI_S$ measures for connecting policy, business, ecological, economic, environmental design, thermodynamic and other scientific views of energy systems is included in the discussion.

The main innovation of the method is a way to use physical causation to locate energy requirements that business information does not record. What is missing from models when describing physical systems comes naturally, in the form of unanswered questions about energy processes, that statistical models don't raise, because of the conservation of energy and other explanatory principles of physics for tracing causal connections. Causal models allow energy uses to be found from their physical processes and natural histories, even when recorded data is not available. For example, the physical energy link between the services of people and the technology they operate can be identified from the tiny amounts of physical energy they exert to operate technology, obtained from their food purchases. It is that energy applied to the buttons and levers of machines using the "know how" and "control" provided by people that operate the business and allow both the machines and people to do their jobs.

From the business manager's view that minute fraction of the food energy that employees consume at home to operate machines at work is vanishingly small. It is still paid for from business revenues and is the essential service provided by employees. It is vanishingly small and insignificant only in quantity, compared to the energy consumed by the machines being controlled. From a physical system view the energy consumed in the environment for the business to obtain those tiny amounts of control energy are its largest energy cost of all. That energy to do its work comes only with employees having the choice of how to spend the rest of their earnings, what they do the work to have. It is part of their pay package, agreed to in exchange for their exerting their minute amounts of smart energy to operate the business. They wouldn't come to work and provide their service without it. Their minute amounts of applied energy, delivering "know how" to make things work, are the highest quality energy source in the world, it seems, and it costs large amounts of energy consumed elsewhere to generate it.

In assessing these hidden energy needs we use a "null hypothesis", that it will be more accurate to initially estimate any cost of business as representing an average energy use per dollar than a zero energy use, as if not counted. The error one way is sure to be infinite and the error the other way is likely to be equally positive and negative for a reasonable sample size. We then look for the available information to refine that initial estimate. Why the energy uses needed to deliver any purchase are likely to be average is also due to how widely and competitively energy is used. Energy is a costly necessary resource at every step of delivering any product or service, has a world price, can substitute for most any other resource and product markets seem to reallocate it to wherever it is most valuable. As the energy needs of business are largely in the services of diverse people and businesses with similarly diverse habits, the energy content of most products is logically going to be closer to average, on average, than greatly above or below. Further study, of course, is clearly needed as well.

*1.3 Background of measuring business energy use*

The common method of measuring the energy used by businesses is based on the ISO 14000 world environmental management standards for assessing the energy consumed by production technology using life cycle assessment (LCA) [15]. LCA collects business information about resource uses for production technologies over their useful life, including their supply chains and eventual disposal, using well defined analytical boundaries to measure their total resource needs and impacts [25, 26]. What is not included are the resource needs for which there are no directly traceable records, such as for having employees and using other business services that determine their own resource consumption choices, and leave no detailed records for the business employing them. Consequently the available data sources do not identify that consumption as associated with the business employing the services that generate it.

The available energy use data is generally recorded and collected according where the energy uses occur, instead of according to what productive activities they serve. As a result it becomes named for economic sectors or types of technology producing the records, rather than the businesses employing the services causing it. Considerable statistical study has been done based on the recorded energy use accounts to identify benefits of technology, links between economic sectors [4, 5, 7, 12, 21] and their relation to growth [1, 3-6, 30, 31]. The data is mostly aggregated by various government agencies, such as the Bureau of Economic Accounting and Census Bureau [8, 9] for the US. The energy uses for steel manufacture are associated with the steel industry, for example, and collected in Input-Output tables by industry group. The energy for the steel going into cars, buses and trains used for business commuters will never show up as an energy cost of hiring employees, though.

What the available data has been most useful for, and LCA is an extension of, is accounting for the performance of individual technologies to optimize energy consumption for production processes. It later became relied on for measuring environmental impacts. LCA starts with adding up directly recorded resource uses and then adds traceable uses in the supply chain trees of contributing production technology. Limits to those trees are set using proxy measures for the tails of supply chain distributions that become uneconomic to individually trace. That serves to effectively "disaggregate" some of the I-O data from national energy use accounts, and assign parts to the service of individual business processes. The first effort to use hybrid accounting to trace and disaggregate "indirect" energy use associated with business products seems to have been by Bullard [32], to then be refined by Treloar [25] and others, leading to the current LCA method standards [26].

As under ISO rules for LCA, those methods counted only the energy costs a business pays for required by "production technology" and not for "production services", considering them as "consumption costs" instead, as follows from naming accounting categories by where the data was recorded. There has been considerable discussion of this over the years, with the economists continuing to separate resource "consumption" costs from "production" costs, whether both controlled (equipment) and uncontrolled (people) equally require consumption to do their productive work. Government statistical categories and models also separate the energy use into categories by where it is used, and treat energy for technology as production and for self-controlled parts of businesses as consumption, as if separate systems.

Even leading systems ecologists such as H. T. Odum [33] analyze and model economic energy uses in the environment by separating the energy consumed by technology labeled "production" and the energy uses needed for employing people and other business services labeled as "consumption". A few others including Costanza and Herendeen (1984) [5, 21] have counted some part of both as environmental costs of business. The study of EROI for US coal, petroleum and nuclear energy reserves by Hall, et. all. [9] is an exception, in both comparing different scales of inclusiveness for assigning energy uses throughout the economy for delivering energy to society, and showing estimates of the energy costs for employing all necessary economic services to result in a scale change of ~500% in the energy accounted for. His EROI3 scale of inclusion also roughly corresponds to our SEA3 scale of inclusion. As to which expenses to consider, we initially assume that any cost a business incurs is probably intended to be in the service of the business. We then assign it a unit energy cost, as further discussed in section 2.1, and seek to verify it.

The usual sticking point in the discussion with economists and others practiced in LCA is the idea that if an employee loses their job they will continue many of the kinds of spending they had when employed. That is, however, argued as a reason to not count the spending of the employed worker as an environmental impact of being employed as well as that of the unemployed worker. We think that argument overlooks both how unemployment spending comes from the savings or government services paid by employed workers, as well as the many large environmental costs of supporting all the other kinds of business services which have individually untraceable resource requirements.

### *1.3 Economic sector intensities and the SEA method*

Figure 1 shows the data from the adjusted I-O table by Costanza [5] arranged to show the relative energy intensities for all US economic sectors in 1963, arranged by the scale of each sector's total energy use. The specific data is quite old in one sense, but the general distribution of intensities and scales is probably similar today. The value of the figure is showing the diversity of technology energy uses compared to income for small sectors, and large sectors all close to average.

If these values were adjusted to distribute the energy costs of employees and other business services to the businesses employing them, the intensities per dollar of output would increase for all producer sectors, with the steel mill including the energy costs of employing steel workers, etc. Their variation would decrease, though, as all came closer to average. That would redistribute the costs of the consumer sectors to producer sectors to account for 100% of energy use for production. A separate table for energy use would be needed to show energy costs by sectors for end consumption.

Until that redistribution is done, using energy intensities by economic sector, as from I-O tables like Costanza's would be misleading as estimates of energy use for whole business costs. Line items in business budgets don't usually correspond to any one business sector, and also not to an average group of business products from any particular sector, which is what I-O tables show. Economic sectors are vast aggregations of different kinds of businesses offering highly varied kinds of products. Business budgets call for particular products or services. Even when an item seems to correspond to an industry group that has only one product, say steel beams, or fuel oil, only the recorded technology energy uses for that sector are included. So for any particular product from any sector, the sector data does not show most of the resource uses those businesses pay for being used. All but the production technology costs are scattered over other sectors.

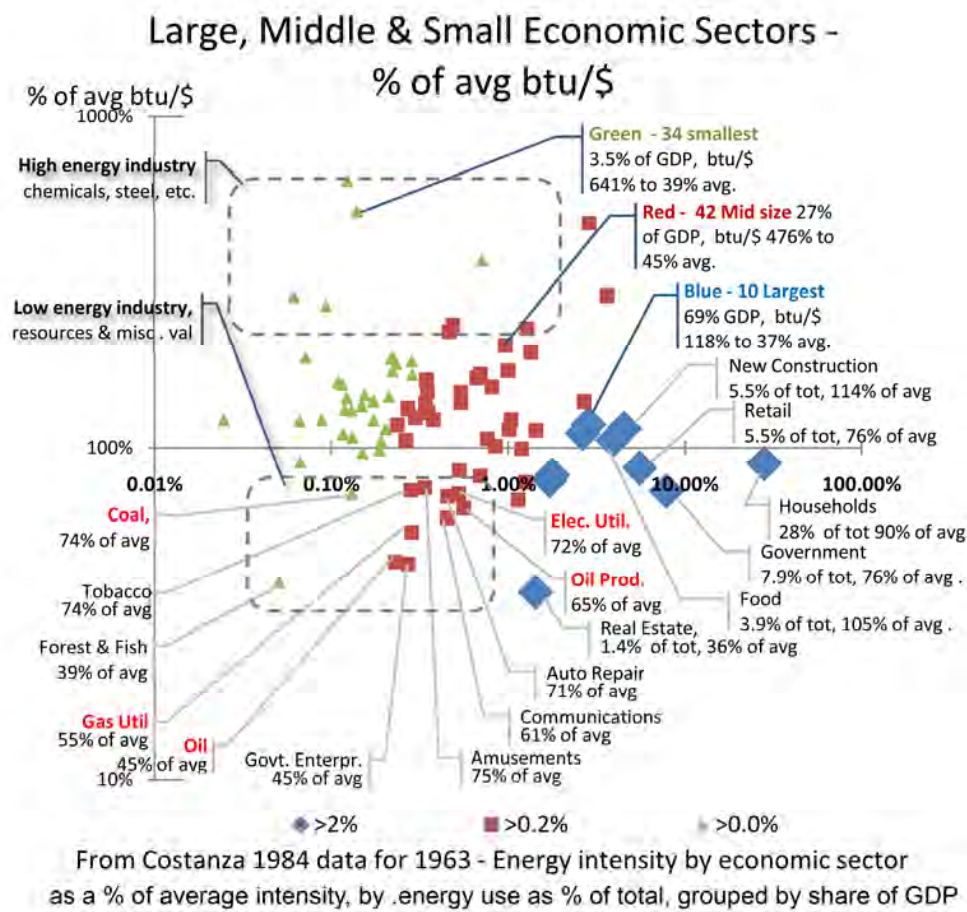

**Figure 1**. US Economic sector energy intensity, btu/$, by sector as a % of the economy (horiz. at center) and by % of average intensity (vert. at lt.) for 1963. The largest sectors have near average intensity, energy sectors below average intensity (for the revenue), and high intensity sectors the most variation. Data from Costanza [5].

If I-O tables were redone to distribute consumption costs to the businesses paying for them, patterns likely to persist are 1) high intensity sectors would still have the most varied energy intensity, 2) the largest sectors would be close to, but below average and 3) the energy producing sectors would have somewhat below average energy intensities. The energy sectors do consume lots of energy, but it also has a high value added, and so the ratio of the energy used to the value produced is low.

Another problem for applying I-O table data to estimating particular energy needs is how the data reflects only national energy accounts. Most products and services have substantial global content. For example, a great deal of the high energy using production for products consumed in the US is now performed overseas, particularly in Asia. EIA data shows energy use in the US beginning to level off starting in the 1970's, even as US GDP and consumption continued to grow [28 fig 3,4]. That rapid divergence between US energy use and GDP is the complete opposite of the consistent world GDP and energy use relationship over the same period. The global trend has been of smoothly growing GDP and GDP/btu [28 fig 1] in constant proportion. Explaining why the global data shows such smooth global trends, and national accounts do not, has been argued as a statistical fluke or the averaging of random variation. Our expectation is that it is a result of the global economy working smoothly, to allocate resources according to the comparative advantage of productive differences for individual business communities around the world, as free market theory has always suggested it should.

### 1.4 Background on EROI for Wind Turbines

Kubiszewski et al. [12] performed a meta-analysis to summarize the net energy of wind turbines based upon a suite of previous studies of 114 calculated values for EROI (see Figure 2). The wide spread of the data shows evidence of large inconsistencies in the methods of defining which energy inputs to count. The variation is over an order of magnitude with reported values from over 50:1 to near 1:1. The average EROI for all studies was reported at 25:1 although the average for operational LCAs (those based upon actual performance of a turbine) was lower at 20:1. There is as yet no national account data for a wind energy sector industry group to compare. Wind utilization estimates varying from 15% to 50% also add to the inconsistency in assumptions presented.

Kubiszewski et al. described process analysis methods, including LCA, and compared them with studies using I-O table data. The former showed an average EROI of 24:1 while the latter had an average EROI of 12:1, a difference some attributed to how process analysis involve a greater degree of subjective decisions [12]. The differences in capacity factors and from omitting the untraceable energy needs of labor and business services required [13, 14] would seem to account for the variation. These results are comparable to the method presented here only as studies of business-scale energy use for which there are no industry group studies.

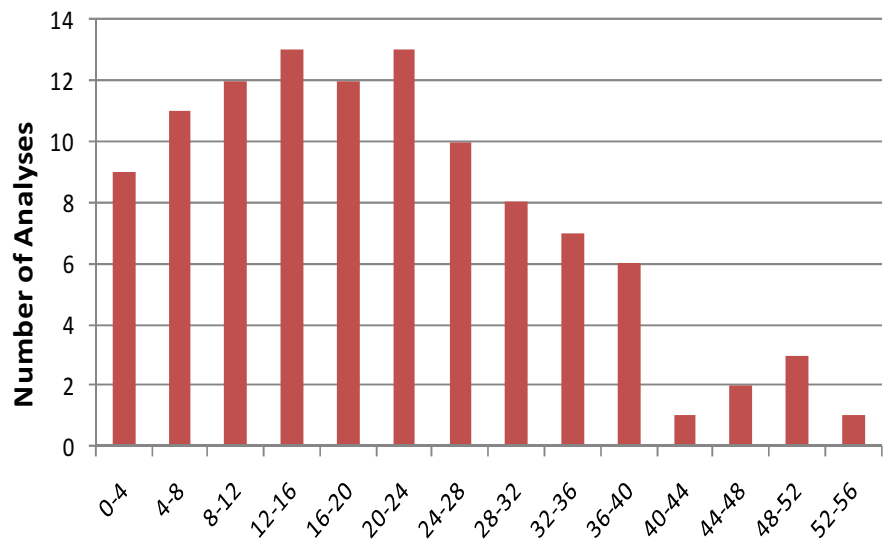

**Figure 2**. The frequency distribution of EROI for wind turbines as studied in [12] shows an unusually wide range, showing the use of inconsistent standards.

## 2. Methods

The SEA method allows us to disaggregate global energy use data for individual business costs. It uses "hybrid" accounting to combine recorded and proxy measures of energy use based on average intensities for monetary costs, not dissimilar to Bullard or Treloar [32. 25]. Instead of using intensities for only technology energy uses, our intensities are for shares of global economic energy use, adjusted in relation to that average if good reason is found.

The key step, from a physical science view, is our method of deciding what energy uses to count. To identify what energy uses are necessary for the operation of a complex environmental system, such as a business, one needs to develop an exhaustive search strategy as a means of deciding what to include. In practical terms that means defining a starting point a then a way to expand the search and then determine when you are at a stopping point. The starting point we use for our case study is an LCA estimate for the life cycle energy costs of the wind turbines and their related capital costs for plant and equipment, for our conceptual model of a Texas wind farm, based on JEDI and VESTAS project data [16, 17]. We could start from any other part of the business too. We just chose to use the usual ending point of business energy assessment as our starting point.

Our procedure is then to ask what else is needed to make those parts of the business work, over and over, until we have exhausted what is needed to deliver the product to market. In that way we let the business as a working system, in its natural form, guide the questioning and determine the limits. That end point identifies the whole working organization of the business as an operating system. A business is a system that makes internal choices for how to work as a whole, operating in a larger open market economy. The economy may determine what its options are, but leaves it a considerable range of choices. As an financial system the line where the internal organization of the business comes to an end, and the external parts and market organization of the economic environment begin, is determined by whether decisions made affecting the business are being paid for from business revenues. Those decisions paid for by the business define the business as a system of complex design organized by the choices of its decision making parts for operating in the environment they face.

A business decides how much to pay workers, depending in part on what kinds of employee skills and self-reliance it needs and what pay package would attract them. A business will generally not determine what city services it gets or pay for them as operating costs, though they will be paid for as essential environmental costs of operation through paying taxes. A business does not pay for the train siding a mile from their plant except in user fees, as that is a service for sale by another business. A business might choose to pay for a community golf outing for its executives to socialize with local business leaders, or support popular political parties, considered as costs of good community relations. A business wouldn't pay for related businesses springing up around it, allowing them to diversify or specialize in higher value added products perhaps, as those are market mechanisms involving decisions by others, though having good community relations might help local industry development of that kind. In one society or another, or for family operated versus publically owned businesses, business decisions may be made very differently. So other criteria may sometimes be needed to distinguish the internal organization of the business from the external organization of its environment. The special task of analysis, that causal models allow and information models don't, is estimating one's lack of information about untraceable energy uses. The search method sends you looking for missing information. This is actually the great benefit of using a physical systems model. An information model would not tell you what information is missing. Using our approach following the working parts of the business by their physical connections then lets us assign energy costs to their dollar value. That is done initially by using the well established consistent relationship between the measures of global purchased energy use and GDP as a basis for equating shares of one with equal shares of the other. That is how we calibrate our "proxy measure", by attributing shares of global energy use in proportion to shares of global economic product. This is where the null hypothesis applies, that average will be more accurate than zero, and the following question is whether other information is available to assign a particular intensity above or below average.

Businesses do not generally pay for things they don't need, so the functional boundary of a business's energy uses would generally match what a business pays for. We did not arrive at that conclusion backwards, by just making an arbitrary choice to start using a different formula for energy estimates. We found that the choices paid for coincided closely with what a business physically needs to independently operate in its environment, by going step by step in accounting for necessary energy uses for which there was no other record. It's the exhaustive search for the parts that need to work together, seeing what organizational unit they are part of and assessing their energy needs, that makes the link of physical causation as good as having a receipt for the energy use.

Our demonstration procedure detailed in Sections 2.1-4 below, is to identify working units of the business ($SEA_N$) to assess and combine. For each we use a table necessary operations, for carefully combining the "technology energy use" ($T_E$), recorded in energy units, with estimated "economic energy use" ($E_E$) recorded in money units following a business plan, so GED is $\Sigma SEA_N = T_E + E_E$ ( adjusted for overlap), assessing predictable energy needs over the lifetime of the investment. We assign economic and technology intensity factors, Tii and Eii in relation to the world average energy intensity $Ei_W$ for translating energy to money or money to energy, using whatever method seems best on a case by case basis. As part of a whole system approach we complete the search with questions about missing information that might remain unaccounted for. We define EROI in the normal way, though by including all energy demands to the point of sale and release of the product for use by others, has new meaning.

What becomes most clear is that understanding the true scale of energy needs of business is well worth the added uncertainty of combining precise data with imprecise econometric measures. It demonstrates that easy and imprecise measures are far more accurate than time consuming precise measures when the latter are a small fraction of the total. As part of a whole system assessment method, the further task is to consider how a better understanding of the business as a system helps you understand its wider roles in the economic and natural environment, that affect business and public values and decisions. We use that discussion for pointing to a sampling of other directions of study.

*2.1 Measurement  strategies for System Energy Assessment*

We use hybrid accounting to combine precise measures of energy use that are very incomplete with rough statistical measures that are very comprehensive (Figure 3). When combining direct records of energy use with statistically estimated energy uses the statistical estimate may need to be reduced to having duplicate recorded and estimated amounts for the recorded energy use (Figure 5). We also develop a strategy for the problem that every fuel use also costs money, and so has both embodied energy content in the economic services that delivered the fuel, in addition to the physical energy bound in the fuel itself. The money paid for fuels is paid to other people for the rights to it, and not for the fuel itself. The fuels themselves come from nature, for "free" and are never paid for except in the energy cost of extraction.

How we proceed with presenting the method is a little more involved than one would need to use in practice. For demonstration we identify six organizational scales, and go step by step asking what else each needs to operate, assessing and combining energy values for traceable technology ($T_E$) and individually untraceable economic services ($E_E$). We start from the LCA estimate for the energy needs of the principle capital investment technology for the wind farm as the first value of $T_E$ that we call $LCA_E$, and consider the smallest whole working unit needed for the wind farm. The technologies of the supply chain businesses are passive equipment that can't operate by itself, though, without the employees and the other business services to operate those businesses. When we add energy use estimates for the active self-managing parts a combination of parts that could operate by itself results, the supply chain, that we call SEA0. Then we ask the same question again.

*Finding boundaries starting from accountable parts*

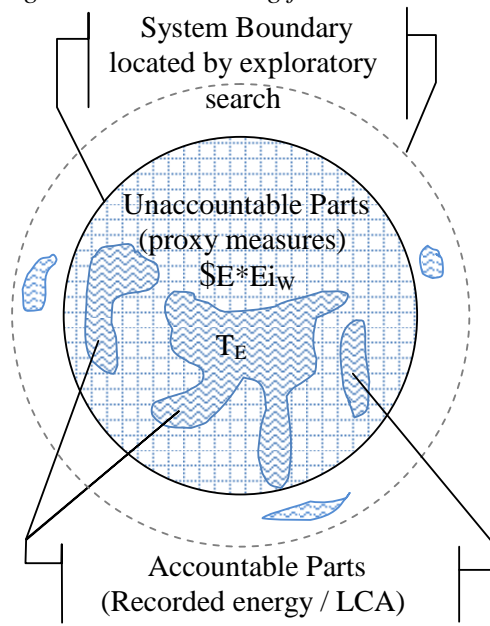

**Figure 3**. Whole system energy use with overlapping direct and proxy measures. Counting recorded business technology energy use ($T_E$) and with economic energy use ($E_E$) at the world average intensity ($\$E*Ei_W$), to correct for overlap when combined.

What we first find is that the principal technology of a wind farm needs a field operation to control and maintain it, and so we add to the accumulating total the needs for those operations and services calling it SEA1. We then do the same to add the operation costs of the wind farm's business office to make SEA2 and the costs of its corporate management to make SEA3. That completes the set of internal units of organization and business costs to account for. We then assess the energy needs that a business will pay for to maintain the external business environment, that we call SEA4, paying for financing costs and taxes to be used by government (Figure 4).

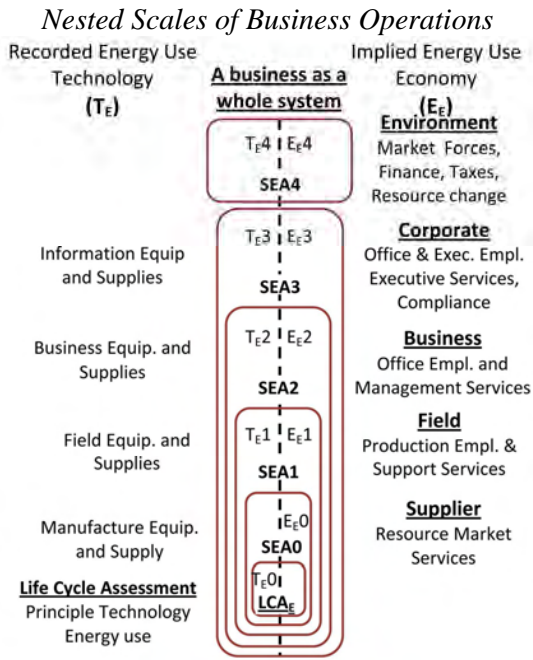

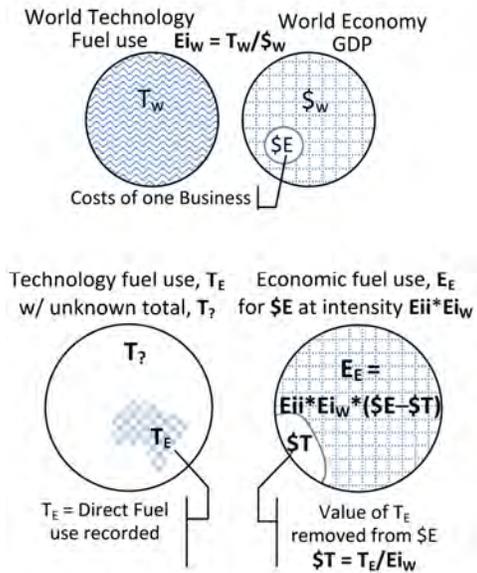

**Figure 4.** System Energy Assessment (SEA) combines technology energy use ($T_E$) with energy uses of needed economic services ($E_E$), using the nested working units to organize the search.

**Figure 5**. If needed to avoid double-counting the value of $T_E$ as $T$ at the world average $Ei_W$, is taken from the item cost $E$ before applying the scaling factor $Eii$ for the proxy measure.

$LCA_E$: the direct energy consumed to deliver and operate a technology measured by LCA

SEA0: adding the indirect energy needed to deliver and operate the technology

SEA1: adding the total energy needed for field operations,

SEA2: adding the total energy needed for managing business,

SEA3: adding the total energy needed for corporate management of the business,

SEA4: adding the total energy for the economic environment, costs of society, taxes, finance.

This series of estimates treats the business as a nested hierarchy of larger scales of organization, with each larger scale serving as the environment for the smaller scale. We present it this way here to illustrate the learning process of asking the same question over and over to locate the boundary of the system as a whole. Each organizational level shown is a "whole business" or "profit center" on its own, then found to also be needing other things to bring its product to market. Economies contain many kinds of nested systems and it is helpful for interpreting results to learn to recognize and describe them. The end point of the search for the necessary parts of the business ends where the product is handed over to someone else. At that exchange the business is paid so it can continue to function. We call the EROI estimates at for SEA3 and SEA4, respectively, "internal" and "external" distinguishing two standard EROI's, labeled $EROI_{Si}$ and $EROI_{Sx}$ if being compared. $EROI_{Si}$ measures of the physical performance of the business independent of the environment it is in. $EROI_{Sx}$ might be used to comparing different business environments for the same business model, and so treating economic development studies as physical systems ecology.

## 2.2 *Double counting corrections*

For each line item in the estimate we can also use either simple or complicated ways of combining values of $T_E$ and $E_E$. For converting financial costs to energy estimates we assign a weight factor (Tii for technology energy and Eii for economic energy) for applying the world average economic energy intensity, $Ei_W$ to the individual cost item. If you were to add up all the purchased fuel uses in the world and combine that with the average fuel use for all end purchases, the total would be exactly twice the total energy use. To correct for that chance of "double counting" when combining $T_E$ and $E_E$ values, there are three options from simple to complex. Option 1 is to ignore the problem for rough estimates when the scale of unrecorded energy uses is evidently much larger that the recorded ones. Then the overlap of combining them will be small compared to the total, and might be within the margin of error for more careful estimates of the total in any event. Option 2 is to carefully choose values of $T_E$ and $E_E$ to not overlap, allowing them to be directly added. That is the case when estimates of $T_E$ are for the total traceable energy needs, like a careful LCA estimate provides, and values of $E_E$ estimate only the business's untraceable energy costs. Then the two can then be added directly without overlap.

Option 3 is a way to begin with estimates of $E_E$ as the combined traceable and untraceable energy uses, and adjust it for partial records of recorded energy use $T_E$, like including actual heating and electric bills  while removing the implied average heating and electric bills in the estimate of the combined total. .  To do that the value of $E_E$ is reduced by the economic value of the recorded energy uses before being combining with $T_E$, removing that implied share of $E_E$ to eliminate the overlap when then combining $T_E$ with the adjusted $E_E$ (Figure 5).   This is also described in equations 7 & 8.  It is interesting and important to note that if lack of information requires assuming the business has average economic intensity, a weight factor Eii = 1, the arithmetic for adjusting $E_E$ to not overlap with $T_E$ will cancel out, and so have no effect.  That  means that in the normal case, where you don't have reason to assume an energy intensity other than average, *it has no effect to count any recorded energy use*.  Only for costs you know for other reasons to have non-average intensity is correcting for overlapping estimates of $T_E$ and $E_E$ needed.  Experience and the availability of published weighting factors will determine what you choose.  These options can be selected line by line if desired, as part of assigning weighting factors to reflect how much above or below the world average energy intensity ($Ei_W$) is to be allowed for.  Table 3 below is organized for option 3 and used for option 2 where appropriate.  Option 1, of course, saves a great deal of time for quick estimates.

## 2.3 *Calculations for System Energy Assessment*

For world energy intensity ($Ei_W$) we use the EIA world marketed energy consumption and global domestic product corrected for purchasing power parity (WMEC/GDP-PPP).  World GDP-PPP was $59,939 billion ($2005) with 472 quads of purchased energy [11] for an average intensity of 7,630 btu/$ or 8050KJ/$ in 2006.  To standardize on electrical energy units we convert kWh:

$$Ei_W\,(2006) = \frac{\dfrac{472e15\;\text{Btu}}{59{,}939e9\;\$2006}}{3{,}410\;\text{Btu/kWh}} = \frac{7{,}630\;\text{Btu}/\$2006}{3{,}410\;\text{Btu/kWh}} = 2.24\;\text{kWh}/\$2006 \qquad (1)$$

Average economic energy intensity varies widely between national accounts[27], but world energy intensity displays remarkably smooth change and proportionality to GDP. The variation between national accounts appears to reflect comparative advantages for specialization in products and services so we choose to use the global average as the default assumption. The world average shows a regular rate of decline of ~1.3%/yr using an exponential fit to the historic data, by Equation (2), where $x$ is the number of years after 2006 [11, 37, Appendix I]. For example, in 2020, at the midpoint of our 20 year project starting in 2010, the current value of $Ei_W$ will have decayed to 1.87 kWh/$2006. Differences in price and value (utility) for different kinds of fuels (e.g. oil, coal, electricity, etc. having different uses and prices per Btu) [3, 6, 8], were not used for shares of the global mix of purchased fuels.

$$Ei_W\,(x) = 2.24 * (1 - .013)^{X} \qquad \text{(kWh/\$ for 2006 + X)} \qquad (2)$$

It is not necessary to account for the business as having nested scales of organization. We do it largely to help demonstrate the method. We aggregate the additional energy identified at each $j^{th}$ working unit level as $d$SEA$_j$, as shown in Equation (3), where $T_{E,k}$ and $E_{E,k}$ are combined after adjustment to for overlap if needed, for '$M$' business costs assessed, and '$j$' is the level of organization considered. The total energy input (SEA$_N$) for the whole business system is the sum of the added energy inputs for each of the '$N$' business units (Equation 4).

$$d\,SEA_j = \sum_{k=0}^{M}\left[T_{E,k} + E_{E,k}\right] \qquad (3)$$

$$SEA_N = LCA_E + \sum_{j=0}^{N} dSEA_j \qquad (4)$$

Values of technology energy (T$_E$) may be obtained either from fuel use records or from fuel or technology costs or budgets ($T) multiplied by the appropriate intensity weight factor Tii and the average energy intensity of money (Ei$_W$), Equation 5. Values of economic energy use (E$_E$) are similarly calculated using economic costs ($E) multiplied by the appropriate intensity weight factor (Eii) and the average energy intensity of money (Ei$_W$), Equation 6. If those values represent partial measures for different things that may overlap the added step of removing an estimated economic value for T$_E$ from $E is needed to eliminate the overlap between the definitions of the two measures (Figure 5), so $E is reduced by $T before T$_E$ and E$_E$ are combined (Equation 7, 8).

$$T_E = \text{recorded fuel use,} \;\; \text{or} \;\; T_E = \$T \cdot Tii \cdot Ei_W \;\; \text{in relation to cost} \qquad (5)$$

$$E_E = \$E \cdot Eii \cdot Ei_W \;\; \text{in relation to cost} \qquad (6)$$

$$\text{Or, with} \;\; \$T = T_E/(Tii \cdot Ei_W)\cdot \qquad (7)$$

$$E_E = Eii \cdot Ei_W \cdot (\$E - \$T) \qquad (8)$$

*2.4 Models and Input values*

We present two separate models 1) using a table with 20 year average costs without discounting and 2) a cash flow model with discounted costs over time. Both models use the same 20 year business plan based on the JEDI budget for a 100MW wind farm [17] as if located in Texas. All money and energy costs are stated as per kW of total generating capacity. LCA data was obtained from the Vestas Onshore 2.0 MW wind turbine study [16], 13,100,000 MJ (3,640,000 kWh). The distribution of energy types for the LCA account is shown in Table 1, showing somewhat more fuel from oil than the world average. Table 2 displays the key business model inputs, including the wind generation capacity factor and other distributions from the DOE 2008 wind report [15]. Our equations for both partial and total EROI and LCOE measures follow the usual standard definitions as in Equations 9 and 10 respectively. A simple version of our Excel model with data and formulas is available [36].

$$EROI_n = E_{out}/E_{in} \text{ for each SEA level} \quad (9)$$

$$LCOE_n = \frac{NPV(\sum C_T + C_E)}{NPV(E_{out})} \propto \frac{\text{partial costs assessed}}{E_{out}} \quad (10)$$

**Table 1.** The quantity of fuel consumed for a Vestas 2.0MW turbine has an energy content of LCAE = 13,100,000 MJ [16] assumed to cost $150,000.

| Fuel/Resource | Energy Consumed (MJ) | Energy Consumed (kWh equiv.) | Fuel cost ($/GJ) |
|---|---|---|---|
| Hard coal | 2,215,252 | 615,348 | $2.34 |
| Crude oil | 6,036,167 | 1,676,713 | $12.23 |
| Lignite (brown coal) | 445,079 | 123,633 | $1.90 |
| Natural Gas | 1,618,058 | 449,461 | $6.21 |
| Nuclear Power | 392,124 | 108,923 | $21.65 |
| Straw | 0 | 0 | $0.95 |
| Wood | 0 | 0 | $0.95 |
| Other Biomass | 57,917 | 16,088 | $0.95 |
| Primary energy from Hydropower | 2,286,239 | 635,067 | $21.65 |
| Primary energy from wind | 37,184 | 10,329 | $0.95 |
| | | TOTAL Cost of fuels ($) = | $147,958 |
| | | Btu/$ of fuel purchase | 83,777 |
| | Btu/$ for fuel purchase : economy average Btu/$ (2010) | | 11.5 |

Section 3.1 presents the cash accounting model shown in Table 3, for 20 year average costs per kW of generating capacity. To represent a realistic consumer market and achieve an annual net revenue of 11% after taxes, we set a market price of $83/MWh. The tax rate on net revenue at SEA4 is set at 36% to approximate the ratio of total US local, state and federal government costs to GDP. Table 3 shows combining the values of $T_E$ and $E_E$ using assigned values of Tii and Eii, based on each item's character and dollar cost by the SEA method. Starting with LCA$_E$ we add estimates for the other scales of business operating units (SEA0, 1, 2, 3 & 4). Values of Tii for technology items are based on the ratio of LCA$_E$/$ of first costs, or as indicated. Values of Eii of 1 are used except as indicated. Column 8 shows typical budgeting ranges estimates for the input costs in column 1. Section 3.2 presents results of a second similar model for comparing the cash and energy flows as they change over time. It shows world average intensity, $Ei_W$, decaying at the recent normal rate of 1.3%/yr with a discount rate of 6% .

**Table 2.** Some SEA input factors are estimated using probability distributions, while most inputs are kept constant at nominal values.

| Input Variable | Units | Value |
|---|---|---|
| Capacity Factor* | % | μ32.6, σ = 6.7 |
| Equipment Cost* | $/kW | μ = 1,433, σ = 125 |
| Balance of Plant Cost* | $/kW | μ = 483, σ = 42 |
| Annual Operation and Maintenance* | $/MWh | Lower bound: 5, Peak = 10, Upper bound = 30 |
| Loan Interest Rate | % | 6.8 |
| Land lease cost | $/turbine | 6,000 |
| Loan amount | % of first costs | 80 |
| Time limit of loan | yrs | 20 |
| Economy inflation rate | % | 3% |
| Marginal federal tax rate | % of annual profit | 35% |

\* From DOE 2008 Annual Wind Technologies Market Report [26]

## 3. Whole system SEA and EROI estimate models and results

### 3.1 Method 1. SEA table and 20 year average costs

Table 3 shows our assessment of the business plan for the wind farm, beginning with the LCA$_E$ energy content of 90.9 kWh/kW and the implied EROI of 31:1 for delivering the 2856 kWh output. From that we ask what else is needed and add energy requirements for successively larger parts of the business operations required to deliver the energy for sale. The accumulative EROI for each level is shown in column 9 and the SEA and EROI results are graphed in Figures 6, 7, 10 and 12.
The basic procedure for each item in the table starts with either column 1 or 3, a dollar cost or an energy cost. We assigned budget range estimates for all inputs as shown in column 8, with accumulative variances at each level shown underlined and graphed as error bars in Figures 7 and 12. The next step is to estimate values for Tii and Eii as above or below average and establish what value of $T (column 4) to remove from $E (column 1) in calculating $E_E$ (column 6) using Equations 5 to 8 for possible overlap. It's important to note, that if it happens that Eii = 1 then the values of $T_E$ cancel out in Equation 8, making the equation for SEA = $T_E + (E_E - T_E)$. That shows that if you don't know much about the energy intensity of any item, finding some partial records of direct energy use to both add and then subtract in equal amounts, does not improve the estimate. So without doing a fairly careful study the simple estimate using Eii = 1 for average energy use per dollar is implied. It also shows the need for empirical studies to develop other guideline intensity factors for Eii and Tii for various common types of businesses and expenses.

**Table 3.** Whole business SEA system energy Input/Output table, arranged by business unit scale

| Output per kW capacity at 32.7% factor | | Value | $/kWh | kWh | oper. | net | AvgCost | Wh/$ | Tax | |
|---|---|---|---|---|---|---|---|---|---|---|
| | **Electricity Sales** | **$236** | **$0.083** | **2,856** | $129.94 | 82.0% | 0.0455 | | | |
| | Average for Economy | | | | | | | 1,883[2] | 36.2% | |
| | | 1 | 2 | 3 | 4 | 5 | 6 | 7 | 8 | 9 |
| | | $E | Tii | $T_E$ | $T | Eii | $E_E$ | SEA | Est. | EROI |
| **Inputs per kW installed capacity / yr** | | Cost | % Avg | kWh | Value | %Avg | kWh | kWh | Range | |
| **LCA$_E$** | **Primary Technology & Equip.** | | 13.1[3] | 90.9 | $3.70 | | | 90.9 | 0.1[5] | 31.42 |
| ***d*SEA0** | annualized Tech & Equip. Cost | $71.63 | | 0.00 | $48.26 | 1.5[4] | 66.03 | 66.0 | 0.15 | |
| | annualized Phys Plant Cost | $24.15 | | 0.00 | $0.00 | 1.5 | 68.22 | 68.2 | 0.15 | |
| | Subtot& Range | 95.78 | | | | | | 134.2 | 0.30[5] | 12.68 |
| ***d*SEA1** | Field technology | $18.12 | 0.5 | 17.2 | $9.13 | 1.5 | 25.4 | 42.6 | 0.2 | |
| | Field fuels | $0.27 | 12.1[3] | 6.15 | $0.00 | 1.0 | 0.51 | 6.66 | 0.2 | |
| | Field Business Services | $0.20[6] | - | | | 0.9 | 0.34 | 0.34 | 0.3 | |
| | Field employees | $2.75 | - | | | 0.9 | 4.66 | 4.66 | 0.3 | |
| | Subtot& Range | 21.34 | | | | | | 54.25 | 0.37 | 10.22 |
| ***d*SEA2** | Business technology | $0.25 | 0.5 | 0.24 | $0.13 | 1.5 | 0.35 | 0.59 | 0.3 | |
| | Business Fuels | $0.54 | 12.1 | 12.29 | $0.00 | 1.0 | 1.02 | 13.31 | 0.3 | |
| | Operating Business services | $1.50 | - | | | 0.9 | 2.54 | 2.54 | 0.3 | |
| | Business salaries | $1.54 | - | | | 0.9 | 2.61 | 2.61 | 0.3 | |
| | Subtot& Range | 3.83 | | | | | | 19.05 | 0.31 | 9.57 |
| ***d*SEA3** | Corporate technology | $0.10 | 0.5 | 0.09 | $0.05 | 1.5 | 0.14 | 0.24 | 0.5 | |
| | Corporate Fuels | $0.05 | 12.1 | 1.14 | $0.00 | 1.0 | 0.09 | 1.23 | 0.5 | |
| | Corporate operations & | $0.50 | - | | | 0.9 | 0.85 | 0.85 | 0.3 | |
| | Invest Land & Local Taxes | $3.00 | | | | 0.9 | 5.08 | 5.08 | 0.2 | |
| | Invest Fees & Insur | $5.34 | | | | 0.9 | 9.04 | 9.04 | 0.2 | |
| | Subtot& Range | 8.99 | | | | | | 16.44 | 0.24 | 9.07[9] |
| ***d*SEA4 0.0** | Finance cost estimate | $69.9 | | | | 1.0 | 131.68 | 131.6 | 0.16 | |
| 0.1 | Cost of Government estimate | $13.2[7] | | | | 0.9 | 22.47 | 22.47 | 0.2 | |
| 0.2 | Production tax credit | -$35.0[8] | | | | 0.0 | 0.00 | 0 | 0 | |
| | Subtot& Range w/o PTC | 83.18 | | | | | | 154.1 | 0.16 | 6.09 |
| | **Project Totals SEA, Range and EROI** | **$213.12** | | | | | | **469.0** | **0.16** | |

*Symbols*: Tii = tech fuel use rate factor, $T_E$ = tech fuel use intensity total, $T = average economic value added for $T_E$, Eii = econ fuel use rate factor, $E_E$ = econ fuel use intensity total, SEA = total energy used, *d*SEA# = change from prior level

1. The value of electricity sales, for a capacity factor of 32% and market price for after tax net revenue of 11%
2. Average economic energy intensity, Ei$_W$ from EIA data = 1.883kWh/$, declining at ~1.24%/yr. over time
3. Tii wt. factor for LCA fuel use, .03, gives the price of LCA measured fuel use in relation to Ei$_W$, for the budgeted fuels it is 12.07, to give the energy value of purchased fuels based on cost of fuel oil, in relation to Ei$_W$.
4. Eii wt. factors assign above or below avg intensities. If all Eii's = 1.0 the table collapses to SEA4 = tot$*Ei$_W$
5. Input Range Estimates, are judgmental estimates for each line item, and underlined to indicate the accumulative range of variance for the total energy accounted for, as seen in Figure 7 bar chart.
6. Cost categories missing from the JEDI model were given estimates.
7. Taxes on net revenue are 36% of net revenue, approximating the ratio of total US local, state and federal government costs to GDP, from an online calculator http://www.usgovernmentrevenue.com/yearrev2008_0.html
8. The production tax credit considered in the financial model is assigned an Eii of zero and not, considered as a transfer payment from other tax payers and not included in the cost totals here or considered as an energy source.
9. The accumulative internal EROI of 9:1 and external EROI of 6:1 indicate the energy available to society before and after including the basic operating costs of the economic environment, *d*SEA4

*3.2 Summary of Method 1. results*

As more of the needed business operations are counted from SEA0 to SEA3, and we count costs further removed from the high cost heart of the business operation, we find a succession of smaller changes in the partial estimates of EROI (Figure 6,7,8). When crossing the boundary from accessing the business's internal to external environment needs, going from SEA3 to SEA4, quite large monetary and implied energy costs and uncertainties are found again, associated with financing, unpredictable net revenues and taxes. As a metaphor, it portrays the business operating as a fairly well defined ship navigating relatively large and unpredictable shifting seas of other things. The partial estimates of EROI decline from 31:1 at $LCA_E$ to 9:1 at SEA3, with an increase of 350 % of in the energy use accounted for. It further declines to 6:1 at SEA4 for an added 150% in the energy use accounted for.

One of the interesting points is how the intensity factor of Eii = 1.5 (Table 3, col. 5, 2nd & 3rd line) was arrived at for the capital costs of the business. In an initial attempt to use realistic values, 1.5 was a test value for using the Option 3 method of avoiding double count. On later examination it was discovered that the same answer would result if we treated $T_E$ and $E_E$ values as entirely separate (using the Option 2 method), using Eii = 0.9 to represent only the untraceable economic energy for the supply chain business operations for the $LCA_E$ technology package. The Eii of 0.9 was estimated by taking the energy use totals from all the economic sectors provided by Costanza [5] that seemed associated with the least traceable energy uses for operating businesses, as a share of the total, around 90%. That avoids the potential of double counting by having a complete estimate of the traceable technology energy use, and an intensity factor scaled to estimate only the untraceable economic energy uses.

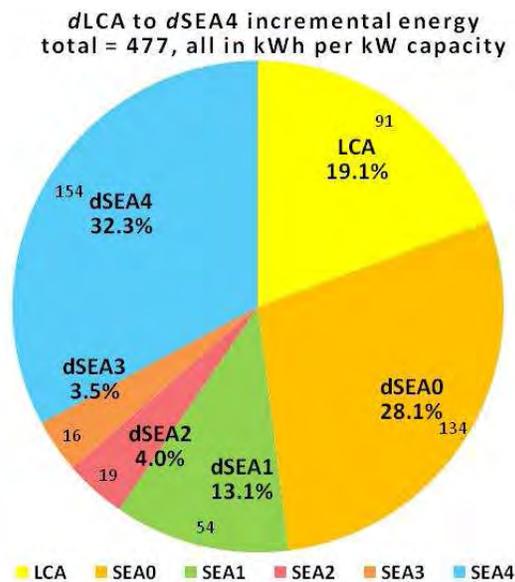

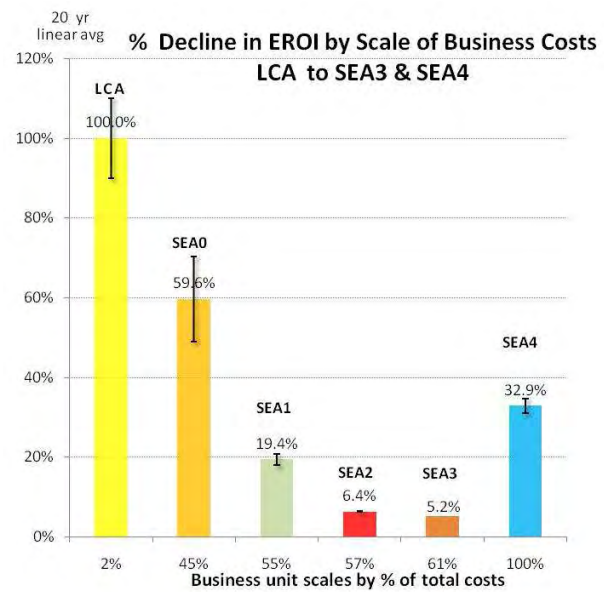

**Figure 6**. Annualized estimates of 20 yr energy use added at each scale of business unit, $LCA_E$ to *d*SEA4. Accounting for all internal costs at *d*SEA3 and adding environmental costs at *d*SEA4

**Figure 7**. Successive effects on value of EROI for counting larger nested scales of business operation, showing declining effect until adding environment costs at SEA4

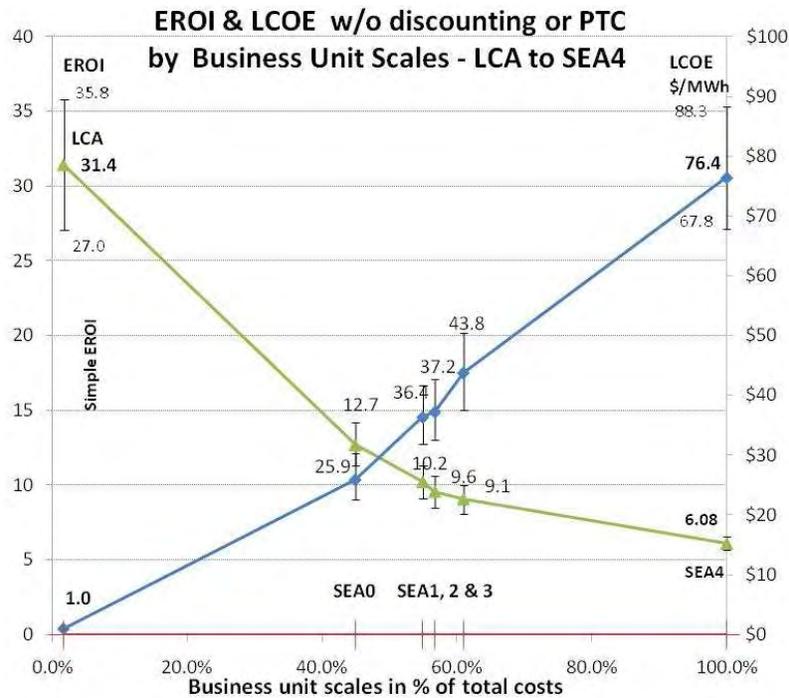

**Figure 8.** Values of EROI decrease and values of LCOE increase as the size of business operation considered increases. Uncertainties assigned in the cash flow model show expected variation in proportion to scale.

Figures 6 and 7 show the fractions of additional energy use accounted for and the corresponding fractional reductions in EROI. The technology energy use (LCA$_E$) accounts for 19.1% of the total, and the value of EROI that starts at 31:1 declines by 81% to 6:1 by SEA4. In Figure 7 you see EROI decreasing by smaller proportional steps from SEA0 to SEA3, but then at SEA4 decreasing by a larger step again, for the large environmental costs of financing and taxes. Figure 8 shows the estimates of the LCOE breakeven price for the electricity, rising from approximately $.002/kWh at LCA$_E$ to $.076/kWh at SEA4. LCOE is shown for the partial cost estimates at each level, just as the partial estimates of EROI were based on partial assessments of the energy needed at each system operating level. We find that the LCOE substantially rises and EROI declines as we consider more of the business operations needed. The curves have different shapes because there is more energy per dollar embedded in the LCA$_E$ and SEA0 business costs.

### 3.3 Cash and energy flow account results

The cash and energy flow model lets us consider the business system as a financial planner would, but from both its dynamic financial model and energy model views. We use a discounted cash flow analysis to analyze monetary flows and for the energy flow model we use a somewhat simplified version of the linear model presented in Table 3. Figures 9 and 10 show the annual and cumulative revenues, respectively, for breakeven operation of the wind energy business. For the LCA$_E$ level almost no cash flow is shown since the revenue required is only to pay for the 90.9/kWh of fuels estimated for the capital costs of one kW of generating capacity, the cost of purchasing the $LCA_E$ estimated fuels to manufacture, install and operate the technology of the turbine.

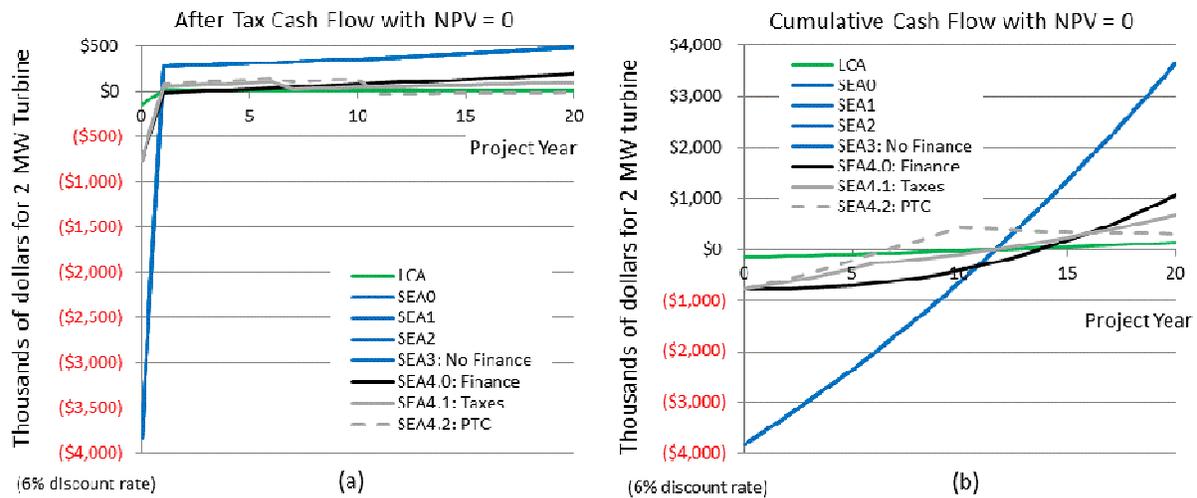

**Figure 9.** Revenue for discounted costs and breakeven pricing. The annual (a) and total (b) after tax (when applicable) cash flows assume a 6% discount rate, showing the high cost effect for first year capital expenditures.

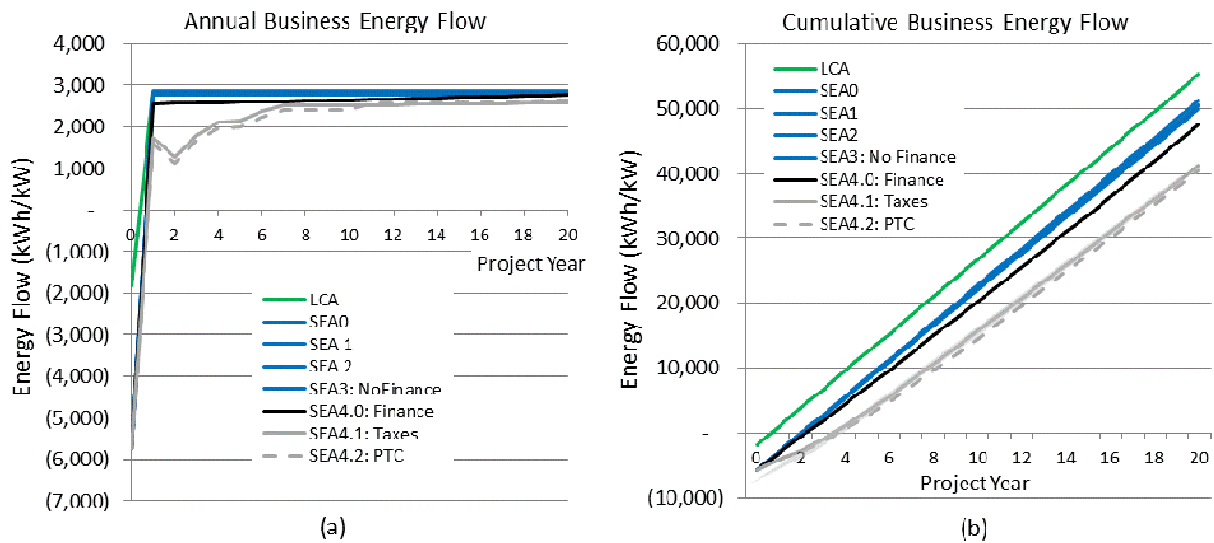

**Figure 10.** The annual (a) and cumulative (b) energy flows (kWh/kW) for each SEA level. The SEA4.2 (PTC) level is shown as if the tax credit was an energy gain to the business, though no energy is saved by it, showing the difference between physical and financial assumptions.

The larger business costs for capital, installation and operating costs at SEA0, 1, 2 & 3 (all shown as the same line in Figure 9 and 10) start very negative for borrowing the capital costs, and turn positive the next year. The net energy flows (Figure 10) also start very negative as the larger costs are accounted for in SEA0, 1, 2 & 3 but due to the assumptions made shows a payback period of only about 1.5 years. The corresponding monetary payback period is almost 12 years. The differing assumptions used for the SEA4.0, 4.1 and 4.2 don't include the initial or operating costs and vary over time. A variety of good questions for how to represent the real energy and cash flows are raised.

By assuming a Production Tax Credit (PTC) subsidy the financial modeling suggests that the revenue gift is also an energy gain for the system, and an implied reduction in the energy needed to produce energy. It shows how standard financial assumptions need to be carefully examined, as physical processes don't change with accounting tricks. The effect is shown in figures 9 & 10 as a dashed line reducing the annual cash flow for the first 10 years. The PTC assumption effectively reduces the investment payback period from nearly 12 years to almost 6 years. Of course, the PTC does not increase wind turbine output, but people not thinking about what they are measuring might confuse the tax credit as an energy grant, and show it as boosting the energy delivered to society.

## 4. Interpretation, application and future work

### 4.1 Whole system comparative value

To compare the energy productivity of our wind farm business model with other businesses, we use the world average GDP produced per unit of energy to define a new benchmark, monetary return on energy invested (MREI), measuring the income produced per unit of energy used (Equation 11).

$$MREI = Revenue / Energy Cost \qquad (11)$$

The world average economic value added for energy use is $1/Ei_W$ (see equation 1 & 2). In Figure 13 bar-2 (100%) & bar-3 (95%) show, respectively, the world average revenue produced for the SEA4 level of energy use compared to the estimated revenue of the wind farm, estimated at SEA4 to for 11% net revenue after costs and taxes. For general comparison bar-1(85%) shows the total expenditures at SEA4 and bar-4 (143%) shows an artificial "retail value" as if the wholesale price were marked up 50%. The wholesale MREI value for the electricity generated by the wind farm seems to be 5% below the world average for using that same amount of energy.

The Costanza data (Figure 1) indicated that businesses in the energy sector generally produce significantly higher than average economic value for the energy used (i.e. lower than average energy intensity for the value added). Some of the model assumptions that might be changed to show the wind farm producing more economic value for the energy are 1) lowering the $Eii$ value of 150% of average for the technology costs, 2) raising the estimated wholesale market price of electricity from $83/MWh to show a greater profit margin than 11%, 3) raising the rate of wind utilization from the estimated capacity factor of 33%, and 4) distributing the capital costs over more years. It might also indicate that the estimated US combined tax rate of 36% on net revenue is higher than the world average. That raises the question of whether societal overhead costs determine what kinds of businesses can prosper in a given society. A variety of societal overhead costs in the US do seem to have been persistently increasing, such as environmental mitigation costs, maintenance costs for increasingly complex urban infrastructure, healthcare, retirement and education, etc.. This one benchmark, then, seems to raise a wide variety of important questions. The accuracy of the measure

itself is likely to be improved with use. What's interesting is seeing all those questions as connected by one number, perhaps indicating this is as good way to look at the interaction of financial, energy and environmental variables.

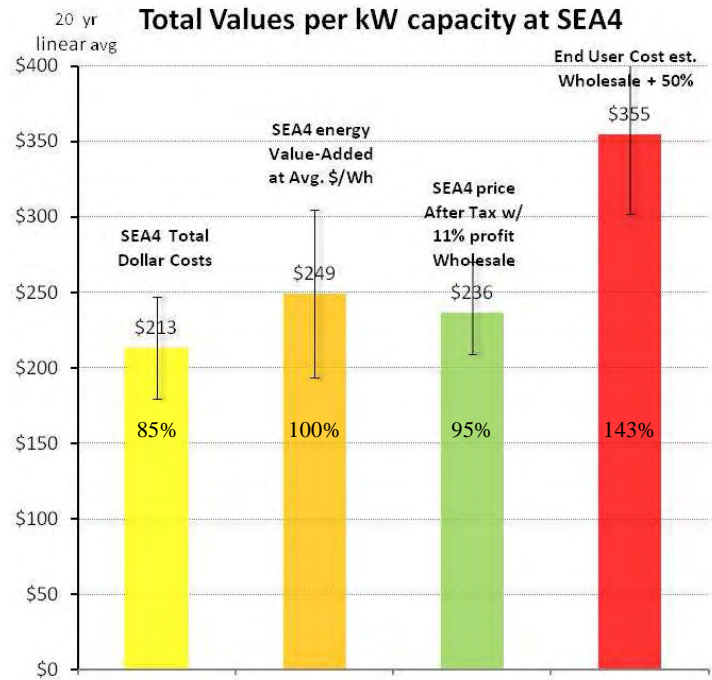

**Figure 13.** Monetary Returns on Energy Invested (MREI) – World Average monetary returns on SEA4 energy use (bar-2 ), Estimated Actual returns (bar-3). A possible whole business energy performance benchmark.

*4.2 Whole environmental assessment*

Life cycle assessments, whether using LCA for technology impacts or SEA for business energy use as whole systems, help people understand how we organize economic activity and interact with our environment. Environments themselves are naturally moving targets, making whole system assessment partly a matter of finding your place in a long story of change. Our historic two hundred year economic experience of accelerating growth, as technology made resources ever cheaper, is now reversing. Increasing resource costs are due to declines in the rate at which new resources are being found, rising competition for them, and also the increasing physical costs and complexity of extraction. Those kinds of reversals in the direction of change are often possible to clearly identify, even from anecdotal information when consistent definitions and patterns of change are lacking. Simply a lag in the rate of finding new reserves is clear evidence of it, as occurred for world oil reserve discoveries in the 1950's [34 fig. 2]. Reversal in the rate of increasing productivity of resource investment is often irreversible, as resource development is generally a comprehensive search, and so indicates natural development limits being faced.

SEA studies provide a way for businesses to physically measure their total exposure to some of these environmental changes. One could also use SEA studies to measure the historical trends in EROI to better identify historical developmental processes that were occurring, and so better project future energy or other resource needs, degrees of environmental resistance and the kinds of business models needed in the future. Business growth strategies change considerably when moving from limitless expanding opportunity to seeking a level of stability, for example. Being able to observe that or other changes in the directions of change in the economic environment could greatly affect business operating conditions and decision making and in the advice given to investors generally.

In the 1750's the difference between heat and temperature was first recognized by Joseph Black, finding that temperature is an energy intensity, not an quantity, but that an intensity could be used to calculate heat as a quantity if it was integrated over a defined volume as a boundary. His friend James Watt learned of the idea and applied it to his work inventing steam engines a decade later. It changed the world. SEA represents a somewhat similar change in measurement science, from thinking about counting visible energy uses by pre-defined accounting category to totaling the functional energy needs for a business as a whole working unit operating in the business environment. It may not create businesses with ever greater energy impact, repeating Watt's magic for creating efficient machines and proliferating energy use. It does define a way to apply the quantitative relationships of energy physics to economic systems, though, and so to apply thermodynamics, the conservation laws, entropy, etc., to business and economic questions. That could change our way of thinking about them, and be very informative about how to make them sustainable.

One immediate use of SEA measures might be for allocating government subsides according to measured performance for delivering sustainable energy to society. It could guide the design and use of tax credits, creating rating systems to help guide intelligent investors or to prioritize research policy goals. It might similarly be used to help allocate tax penalties, to more fairly distribute societal costs of eliminating $CO_2$ pollution in response to climate change or to facilitate other resource depletion policies. One policy objective might be to restrict the use of supplies of presently cheap but increasingly costly diminishing resources, reserving their use for making higher cost but sustainable resources usable.

That EROI depends on system overhead costs should be an "eye opener". The sustainability of a society built for cheap energy and low overhead is brought into question if its unproductive overhead costs steadily increase as its resources become progressively more expensive. That tipping point can be approached by being simply unresponsive, by "doing nothing". Just continuing to make desirable but unproductive luxuries and infrastructure essential for societal functions, until maintaining it becomes unaffordable, a path of retreat may not then exist. The original study of this relationship, the examination the EROI an advanced technological society must have to operate, was by Hall et. all. [2], and is being revised for inclusion in this volume. Using empirical methods like SEA, which rely on using the organization of systems in their natural form to define their physical measures, could add considerable validity and precision to such estimates. That could be quite valuable for making the informed choices about the increasing complexities of the world we need to better understand [24].

*4.3 Pros and Cons of SEA methodology*

It's surprising that relatively easy rough estimates turn out to be more accurate than time consuming efforts relying on precise measures. In this study, after all the effort we went to, the best total value we found for the energy needed for the wind farm differs by 500% (Equation 12) from the method serving as the world standard for estimating the same thing. That total, though, is surprisingly only 15% different from using the simplest possible method of making the same calculation, assigning the global average energy intensity of money to the total project cost (Equation 13). The great effort we went to did change the result much from what we could have assumed from the start.

$$EROI_{LCA} / EROI_S = 31.42/6.09 = 516\% \quad \text{more than ISO standard} \quad (12)$$

$$(\$Etot * E_W) / SEA4 = (\$213 * 1.87) \text{ kW} / 465 \text{kW} = 84.5\% \text{ or } 15.5\% \text{ less than average} \quad (13)$$

It clearly shows that using money to measure the real scale of economic energy use can be both less precise and much more accurate, than carefully counting up traceable energy uses. The great majority of economic energy use comes from the delivery of unreported services scattered all over the whole economic system making them untraceable. That the change so great it indicates not asking the right question, what is called a Type III error. In part it implies that the energy use estimation procedure should be reversed, beginning with the easier but more inclusive method to start. Easy preliminary estimates based on econometric measures would be supplemented with accounts for traceable energy uses affordable. As traceable energy uses are such a small part of the total, the amount of effort to identify them in most cases would have rapidly diminishing returns for altering the total.

That average money uses will have average energy content, and for lacking any better information about most money uses and their energy content, results in our needing to begin relying on money as a measure of energy use. Even if it will take time to understand it, the implication is that money is actually a real form of energy currency, both physically and for our information models, a sort of surprising result. Money is considered as the economy's universal resource. That on average it seems to be a direct measure of our use of nature's universal resource too, indicates that energy is the main resource used to deliver what people value. It seems to make the question of how much energy is used be the same as asking how much of the economy is used. That the scale of money use generally reflects the scale of energy use also clearly implies that our common perception of "decoupling" between money and the environment is a complete illusion. From seeing where the missing energy uses were found, in the branching trees of services hidden from view, the belief that money has no energy content evidently comes from our trusting our lack of information about it.

Going to the extra effort to determine how much above or below average your economic energy use may be will still be warranted in lots of cases, though, like if you have a carbon pollution tax bill to calculate, for example. Production engineers will also still get significant value from understanding the real costs of their own production technology choices too, of course, as obtained by careful LCA studies. Developers and investors will also still want to know as much as they can about things like how much their investments expose them to threatened resources.

How SEA appears to get the scale right is by 1) having a reproducible and improvable way to define the full extent of what is to be measured, using the natural boundaries of the system measured, 2) relying on the world economy to uniformly allocate energy uses by setting a world price for the available supply, and so providing a way to start estimates of energy use for items in a budget, and 3) having the scale of missing and untraceable information so very large it would be unimaginable to get a better estimate the traditional way.  Input-output models [5, 7, 25, 32] as illustrated in Figure 1 show widely varying energy intensity, and so a need to develop ways to assign a variety of different intensities to different things to improve the SEA method.  Except as a share of the whole economy's impacts, indicated by dollar value, it's quite hard to say what impacts any product would have the way "green" products often labeled as having less impact but costing more to deliver.  It's commonly believed they would be the products produced locally, and buying local would help steer the economy away from relying on non-renewable energy.  How the SEA method advances those interests is by allowing a better understanding of whole system energy costs, exposing a business to higher or lower opportunity costs or environmental risks to itself or others.  In the economy of the present, though, local products are often more expensive and so implicitly more energy intensive, for not having economies of scale and specialized content.  Until those kinds of differences are better understood, absent other information, "about average" is a better estimate for the implied energy use of spending than "zero".  As much as prompting a search for better ways to estimate the totals, it also suggests that the general subject of complex organization in environmental systems, that few people seem even aware of, is important for general public discussion.

We think the SEA method makes a useful contribution to defining quantitative physical measures for business systems in general.  It defines a business by the organization of its working parts, considered at a stage of its development in a changing environment, rather than as columns of figures arranged in functionally unrelated categories only leading to a "bottom line".  By using a method of exhaustive search using a repeated neutral question about what's missing, the SEA method empirically locates the functional boundary of whole systems.  That's a possible model for how to calibrate reproducible physical measures of various kinds of distributed systems, using their own organization as a boundary.  The new approach makes such measures comparable and provides a sound basis for evaluating alternate business model choices.  Thus SEA has distinct advantages relative to alternative methods of analyzing the energy inputs and outputs associated with complex environmental systems.

*4.4 Future Work*

The method as presented provides a fairly simple common standard measuring business energy use, but it will take some time and effort to adapt it for widespread practice.  Various industries might discover different ways it needs to be applied to their needs, and large scale econometric studies of different kinds are clearly needed as well.  Questions like how to compare the utility of different fuels would need to be addressed.  There might be a useful relation between the world average GDP/kWh we have used to measure economic value of energy, and the ecological concept of "emergy" in ecological systems, for example.

The mismatch between when wind and solar energy are available and when they are needed seems likely to require analysis of industrial systems of larger scale than individual businesses, for example.

Lots of development problems are like that, needing to apply to unique industrial and natural environments, and to current stages of technological and human cultural development. In some regions local wind energy integration solutions have emerged. For example, Denmark uses pumped hydropower within Scandinavia for storage of excess electricity and exports to other markets. In the Texas grid (Electric Reliability Council of Texas), 4.9% of the electricity in 2008 was from wind power, and the large capacity of natural gas generators on the grid has thus far enabled relatively easy integration of wind. However transmission constraints have restricted wind power flow at many times to lower the capacity factor by up to 10%. Eventually at very high penetrations for wind energy (over 20% of total electricity), newer chemical or thermal battery systems may need to be employed. However, installation of natural gas combined cycle systems may serve the need to mitigate the intermittency of wind at the cheapest cost. Thus, if the energy inputs and/or EROI of each component added to the electric grid is known, one can estimate the EROI of the supply system as a whole for matching the demand.

One strategy for increasing average wind output is evident in the high energy cost of all the initial development costs that scheduled to be repeated every 20 years. Further study would compare different replacement rates and net returns. We might changing from assuming discount rates for the value of money to study business models with a built in cost of savings, to become self-sustaining and self-financing over time. This relates to the long discussion of whether to invest in short or long term development, sometimes in terms of economic arguments over what discount rate to assume in cost-benefit analysis [23]. Those economists choosing a low discount rate tend to find a net benefit for investing in avoiding ballooning long term future costs, such as climate mitigation and resource depletion. Those choosing to make business decisions assuming high discount rates find less benefit in long term investments, but their businesses might be less likely to be sustainable too.

Another area of research needing attention is the basic relation between financial information and the physical economy. It's remarkable that world GDP so closely tracks world energy use, and shows so little sign of sudden shifts in direction. That is not the case for historic changes in asset values, though, demonstrating a tendency for financial markets to develop great bubbles of misinformation about future economic performance. One can clearly see the difference between grounded and ungrounded economic measures in the way the valuation of the US stock market wanders all over and the physical value of the economies has changed relatively smoothly over time. For forty years world energy use has also followed smooth curves in proportion to world GDP. While US GDP and energy use have followed different trends over the period they have also been self-consistent. The US stock market has not followed any of the physical economy trends, though, but seemingly wanders by itself [28 -Figure 1, 4, 5].

Since economic measures that closely track energy use and move independent of it seem common, understanding why different markets do and do not provide reliable whole system measures is important for having confidence in using money as a physical measure. Lots of budget items like financing, profit projections, tax rates, subsidies, returns to investors or discounted values might all introduce speculative information to distort a physical system assessment. Our approach to avoiding the misuse of money as a physical measure was to be careful in assessing individual cost items. For estimating tax burdens we used shares of the total cost of government in relation to the national economy (rather than special rates for special purposes). For profit rates we used a generic rather than a theoretical profit requirement. At least as important as these adjustments to fit the method to the problem seems to be to always treat economic measures of energy loosely, perhaps true to scale but read as if calculated to one significant digit rather than two or three.

Also needing further study, of course, is the real meaning of "average" as it applies to the embodied energy of money, and of how to tell what kinds of spending are above and below average. We've assumed world energy use per dollar to be uniform, so studies of the non-uniformities are needed. Among other ways to study that is a network analysis of how money circulates. For example, if in a month a person gives money to 200 different businesses, and each business receiving the income gives money to 200 different people, then in three months there are 200^3*200^3 partial recipients of any dollar spent. That cascade results in 6.4*10^13 potential end recipients in three months. If you assume there are 5 billion economically active people on earth, each one might receive part of a single dollar spent three months earlier by an average of 13 thousand different paths! It's confusing math, but may be fundamentally important for understanding how our economies work, and how to measure them. A network analysis examining the "degrees of separation" between energy uses communities that add more or less value to energy, and add to the understanding of product space relationships and community development pathways [35]. Finding how to reducing energy use without reducing comfort is probably not done just by just changing one's spending from one thing to another. It's more likely done by earning less but spending on things of more value, a cultural change.

## 5. The scientific questions

The traditional method of measuring the environmental impacts of businesses appears to count only the resource needs businesses record in the process of controlling machines and equipment that are not self-managing. That has left uncounted the resource needs for supporting employees, managers and the services they use to operate business. We showed one way to solve that accounting problem, but it identified such a large discrepancy in results, an increase to five times the original estimate, that it suggests there is something wrong with how the question was asked, a Type III error. In part that prompted a change in our perception of money as a physical measure of environmental impacts, as discussed in section 4.3. It also prompts a question about how common it is that scientific models are based on available information, instead of assessing the working processes said to be depicted.

That both professionals and the public appear to be quite unaware of the scale of energy use required for common purchases. Simple conversions of the ratios of world GDP, energy use and $CO_2$ show that average expenditures like a $6 glass of wine consume the equivalent of 6 times the volume of gasoline in energy and produce around 16 times its weight in $CO_2$. The decimal points matter less than the wholly unappreciated scale. It exposes how our intuitions, and scientific explanations, rely so heavily on the information at our disposal, and how that hides the dimensions and behavior of the environmental systems we are part of. The implication is that we have a great need to begin getting our information about whole system effects in some new ways, and other kinds of environmental models may contain the same error of accounting for environmental processes as controlled by where we find records of them, rather than by how the animating and inanimate parts need to work together.

The wider implications of changing any widely accepted scientific method or its use for guiding world environmental policy, are well beyond what can be addressed here. So too are the wider implications of looking for practical methods for defining measures to fit the form of the thing being measured and checking to see what common measures do and do not. The origin of the discovery, though, is interesting. It seems to be a result of pursuing an accounting problem that required finding a way to treat complex units of organization in the environment as physical subjects, sufficiently well defined to physically measure. Most of the extensive branching trees of energy use needed to bring products to market are hidden from view, and so have gone unobserved and unrecorded. That was the problem that made the difference, identifying a classic "fat tail" distribution that was not visible from the available data. So from the view of information models those energy uses disappeared, until we considered the physical causations involved as "receipts" for the real costs of the services provided and had the luck of finding what seems like a good way of estimating them.

Part of the reason to mention these complex issues, but also keep the discussion short, is that people are accustomed to thinking of physical processes in the environment as following formulas, and they don't. Our cultural awareness of how complexly organized natural systems work is very undeveloped. The systems of nature seem much better described as local developmental processes, having parts that change everywhere at once, following their own emerging dynamics as they respond to local conditions. Environmental systems mostly have actively adaptive parts, and their collective behaviors reflect how new directions of contagious development emerge, first opening up ever greater and then less opportunity for themselves. Human interests and inventions are like that, "stormy", but then even the weather is too. So they only appear to follow formulas when their ways of changing are steady for a while, but can also change direction fairly quickly sometimes with little notice for those not knowing what to watch for. For our measures and theories to fit nature better we would need to both shift our focus from considering natural systems by where the information collects, their "symptoms", to assessing their working processes of "development" that show how they work, while also of course keeping watch for how their regular behaviors may quickly change direction.

## 6. Conclusions

Methods for determining the energy costs of business and the energy needed to produce energy, EROI, have suffered from the difficulty of defining what to measure. The standard LCA method is well defined for assessing the traceable energy costs of business, associated with technology. Most of the energy costs of businesses are not readily traced so LCA has not counted them. SEA corrects that using a combination of "bottom up" and "top down" approaches for assigning shares of world energy use, based on shares of world economic product. That revealed a large discrepancy between the two methods, and many categories of instrumental energy use that were going uncounted.

Using SEA and starting with data categorized by where the records were found, we defined a method of tracing causal links and reassigning energy uses found elsewhere, to "disaggregate" the original categories and reconstruct the functional energy needs of the business operations paying for them. Combining the two kinds of energy information required combining precise but incomplete measures with imprecise but comprehensive measures. We used hybrid accounting to do that, not unlike that used for LCA, but asking a different question. Instead of identifying the task as collecting information from predefined categories, we identified the task as objectively defining a whole system of connected operations so they could be physically measured.

To remove the subjectivity of defining what to measure we asked the question of what else is needed and followed the physical causations to let the natural definition of the system determine its own boundary. It identified the business as a whole system of controlled and controlling parts with a functional boundary of energy needs. That is what located the missing information to account for. We used that reconstructed network of functional energy requirements to define both an energy measure of the business and the physical extent of its working parts, identifying the business as an organizational unit of the environment in its natural form. Thus the method of determining its total energy use also uniquely identifies the individual complex system using the energy, allowing it to be referred to elsewhere as well.

The method that results is straight forward, well defined and can be improved. We presented it with some repetition to demonstrate the search strategy it is based on and discuss different views. We hope people take the obvious shortcuts possible but also look for where the short cuts leave things out and a search strategy or reference notes on what may be missing could be added. In the end the SEA method is not a fully defined procedure, but describes a learning process for discovering the full extent of the working parts of a business as an environmental system. Basing the measure on locating that as the functional boundary of the business, for assessing everything within it, is what turns the qualitative measures for the separate parts into a quantitative physical measure of the business as a whole, for its own boundary.

We feel we have demonstrated a more accurate way to measure and understand the real scope of the energy costs of business, consumer and development choices. We also see it as a framework for doing original scientific research on locally organized systems. It's a procedure for defining a complex system by its boundary, starting a search from one instrumental point, that provides a view of its separate worlds of interior and exterior relationships. It is quite tentative, of course, but the hope is that a way to define businesses as measurable physical systems making it possible for the various sciences to independently study the same subjects, will help us discover the real relationships between money choices and the environment at this time of such great need. Consistent with the view that a business is a component of a complex natural world and the general desire to better understand both, perhaps a way to see its systems as whole individual objects now makes that more practical.

## Acknowledgements

Even a long hard uphill push against what seems like resistance from every quarter has its angels of mercy and good luck to be grateful for. We'd like to thank all the prior contributors to the EROI discussion, especially Charlie Hall, for forging the way and making a subject so critical to the sustainability of the earth's resources acceptable to discuss in "polite society". As co-authors, we also wish to thank each other too and our institutional support, for forming a solid team despite having different approaches, keeping each other honest and pushing each other to maintain high standards. What we most have to thank is the accident of our own passion for the ideas that made the years of effort that led to this paper something compelling and worth the sacrifices and the time.

————————————

## Nomenclature

SEA or $E_{IN}$ – total system energy use

$d$SEA - change from prior SEA level

$Ei_w$ - world average energy intensity per dollar GDP

$\$E_{jk}$ - business costs of kth item in the jth business unit

$E_{Ejk}$ - economic energy of kth item of jth business unit

$Eii_{ij}$ - relative economic fuel use intensity factor

$\$T_{jk}$ - energy use value of kth item of jth business unit

$T_{Ejk}$ - tech energy of kth item in the jth business unit

$Tii_{ij}$ - relative technology fuel use intensity factor

$E_{out}$ - energy produced

EROI - energy return on energy invested

LCOE - levelized cost of electricity

MREI – monetary return on energy invested

LCA - life cycle assessment

$LCA_E$ - life cycle assessment energy

TEA - total environmental assessment

PTC - production tax credit

NPV - net present value

IRRe – internal rate of return for energy

## Abbreviations

| wt | weight | avg | average | invest | investment |
|---|---|---|---|---|---|
| est | estimate | yr | year | insur | insurance |
| tot | total | equip | equipment | tech | technology |
| val | value | | | | |

Appendix I   - World Economic Energy Use

### World GDP PPP, Energy and Energy intensity

| Date | World Quad btu TPES | GDP PPP Billion 2000$ | Intensity Kbtu/$ PPP | Intensity kWh/$ PPP |
|---|---|---|---|---|
| 1971 | 231,081 | 17,540 | 13.17 | 3.86 |
| 1972 | 242,333 | 18,463 | 13.13 | 3.85 |
| 1973 | 255,371 | 19,705 | 12.96 | 3.80 |
| 1974 | 256,863 | 20,251 | 12.68 | 3.72 |
| 1975 | 258,671 | 20,644 | 12.53 | 3.67 |
| 1976 | 272,856 | 21,709 | 12.57 | 3.69 |
| 1977 | 282,560 | 22,642 | 12.48 | 3.66 |
| 1978 | 293,662 | 23,648 | 12.42 | 3.64 |
| 1979 | 302,742 | 24,571 | 12.32 | 3.61 |
| 1980 | 301,883 | 25,098 | 12.03 | 3.53 |
| 1981 | 299,742 | 25,532 | 11.74 | 3.44 |
| 1982 | 299,878 | 25,753 | 11.64 | 3.41 |
| 1983 | 303,257 | 26,542 | 11.43 | 3.35 |
| 1984 | 315,165 | 27,702 | 11.38 | 3.34 |
| 1985 | 323,746 | 28,669 | 11.29 | 3.31 |
| 1986 | 330,516 | 29,692 | 11.13 | 3.26 |
| 1987 | 342,614 | 30,757 | 11.14 | 3.27 |
| 1988 | 354,538 | 32,094 | 11.05 | 3.24 |
| 1989 | 360,368 | 33,254 | 10.84 | 3.18 |
| 1990 | 366,530 | 33,357 | 10.99 | 3.22 |
| 1991 | 370,119 | 33,815 | 10.95 | 3.21 |
| 1992 | 370,045 | 34,556 | 10.71 | 3.14 |
| 1993 | 373,286 | 35,314 | 10.57 | 3.10 |
| 1994 | 376,420 | 36,555 | 10.30 | 3.02 |
| 1995 | 385,904 | 37,830 | 10.20 | 2.99 |
| 1996 | 395,980 | 39,354 | 10.06 | 2.95 |
| 1997 | 399,490 | 40,993 | 9.75 | 2.86 |
| 1998 | 401,860 | 42,085 | 9.55 | 2.80 |
| 1999 | 409,775 | 43,674 | 9.38 | 2.75 |
| 2000 | 418,625 | 45,761 | 9.15 | 2.68 |
| 2001 | 420,069 | 46,940 | 8.95 | 2.62 |
| 2002 | 429,289 | 48,349 | 8.88 | 2.60 |
| 2003 | 444,602 | 50,267 | 8.84 | 2.59 |
| 2004 | 465,820 | 52,884 | 8.81 | 2.58 |
| 2005 | 477,336 | 55,438 | 8.61 | 2.52 |
| 2006 | 489,893 | 58,466 | 8.38 | 2.46 |
| 2007 | 502,920 | 61,748 | 8.14 | 2.39 |
| 2008 | 512,286 | 63,866 | 8.02 | 2.35 |

To characterize the world economic energy intensity it is helpful to show the raw data  and ratios for world GDP (PPP) and TPES purchased energy use from the IEA[37].  Similar world data is available from the US EIA[11] from 1980, showing the same average decay rate of energy intensity over time of 1.3%/yr

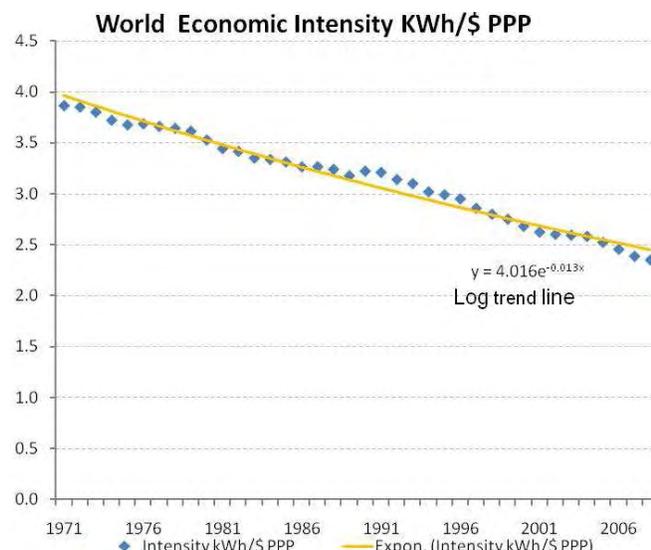

**Figure 14.** Decay rate trend (~1.3%/yr) for world economic intensity (Wh/$) for purchased energy use [37]

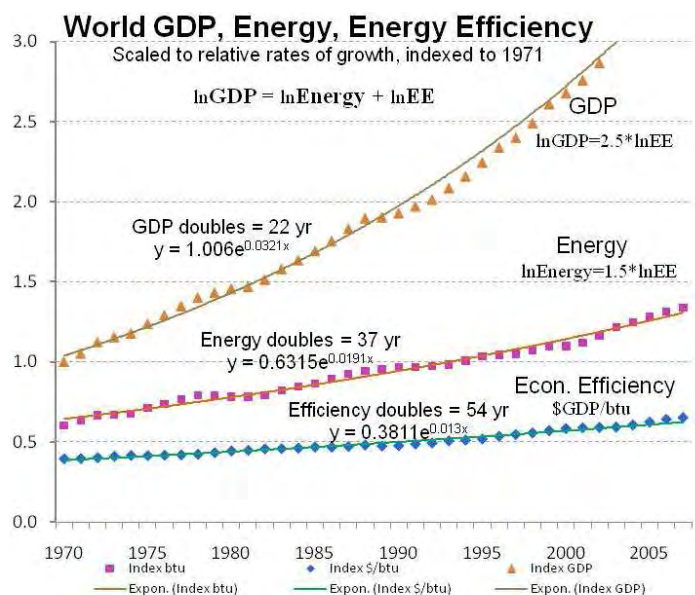

**Figure 15.** Relative historic economic growth trends for GDP, Energy use and Energy Efficiency ($/Wh) four purchased energy use [37]